\documentclass[traditabstract]{aa} 
\usepackage{txfonts}
\usepackage{graphicx}
\usepackage{longtable}
\usepackage{lscape}

\begin{document}

\title{The population of hot subdwarf stars studied with Gaia}
\subtitle{I. The catalogue of known hot subdwarf stars}

\author{S.~Geier \inst{1,2,3}
   \and R.~H.~\O stensen \inst{4}
   \and P.~Nemeth \inst{3}
   \and N.~P.~Gentile Fusillo \inst{2}
   \and B.~T.~G\"ansicke \inst{2}
   \and J.~H.~Telting \inst{5}
   \and E.~M.~Green \inst{6}
   \and J.~Schaffenroth \inst{3}}

\offprints{S.\,Geier,\\ \email{geier@astro.uni-tuebingen.de}}

\institute{Institute for Astronomy and Astrophysics, Kepler Center for Astro and Particle Physics, Eberhard Karls University, Sand 1, D 72076 T\"ubingen, Germany
\and Department of Physics, University of Warwick, Conventry CV4 7AL, UK
\and Dr.~Karl~Remeis-Observatory \& ECAP, Astronomical Institute, Friedrich-Alexander University Erlangen-Nuremberg, Sternwartstr.~7, D 96049 Bamberg, Germany
\and Department of Physics, Astronomy, and Materials Science, Missouri State University, Springfield, MO 65804, USA
\and Nordic Optical Telescope, Rambla Jos\'e Ana Fern\'andez P\'erez 7, E-38711 Brena Baja, Spain
\and Steward Observatory, University of Arizona, 933 North Cherry Avenue, Tucson, AZ 85721, USA}

\date{Received \ Accepted}

\abstract{In preparation for the upcoming all-sky data releases of the Gaia mission we compiled a catalogue of known hot subdwarf stars and candidates drawn from the literature and yet unpublished databases. The catalogue contains 5613 unique sources and provides multi-band photometry from the ultraviolet to the far infrared, ground based proper motions, classifications based on spectroscopy and colours, published atmospheric parameters, radial velocities and light curve variability information. Using several different techniques we removed outliers and misclassified objects. By matching this catalogue with astrometric and photometric data from the Gaia mission, we will develop selection criteria to construct a homogeneous, magnitude-limited all-sky catalogue of hot subdwarf stars based on Gaia data.

\keywords{stars: subdwarfs -- stars: horizontal branch -- catalogs}}

\maketitle

\section{Introduction \label{sec:intro}}

Hot subdwarf stars (sdO/Bs) have spectra similar to main sequence O/B stars, but they are subluminous and more compact. The formation and evolution of those objects is still unclear. In the Hertzsprung-Russell diagram those stars are located at the blueward extension of the Horizontal Branch (HB), the so called Extreme or Extended Horizontal Branch (EHB, Heber et al. \cite{heber86}) and are therefore considered to be core helium-burning stars. 

To end up on the EHB, stars have to lose almost their entire hydrogen envelopes in the red-giant phase most  likely via binary mass transfer. Hot subdwarfs turned out to be important objects to study close binary interactions and their companions can be planets, brown dwarfs, all kinds of main sequence stars, white dwarfs, and maybe even neutron stars or black holes. Hot subdwarf binaries with massive white dwarf companions are candidates for the progenitors of type Ia supernovae. They are possibly ejected by such supernovae as hypervelocity stars (see Geier \cite{geier15}). Hot subdwarfs dominate old stellar populations in blue and ultraviolet bands. Their atmospheres are peculiar and can be used to study diffusion processes like gravitational settling or radiative levitation. Furthermore, several types of pulsating sdO/Bs have been found and turned out to be well suited for asteroseismic analyses. For a comprehensive review about the state-of-the-art of hot subdwarf research see Heber (\cite{heber16}).

SdO/B stars have initially been found in surveys for faint blue stars at high Galactic latitudes (Humason \& Zwicky \cite{humason47}). The first larger area surveys for such objects were the Tonantzintla survey (TON, Iriarte \& Chavira \cite{iriarte57}; Chavira \cite{chavira58,chavira59}), the Palomar Haro Luyten survey (PHL, Haro \& Luyten \cite{haro62}), and the Palomar-Green (PG) survey (Green et al. \cite{green86}). The Kitt Peak-Downes (KPD) survey covered a substantial area in the Galactic plane for the first time (Downes \cite{downes86}). Collecting these early discoveries Kilkenny et al. (\cite{kilkenny88}) published the first catalogue of spectroscopically identified hot subdwarf stars. This catalogue contained photometry, spectral types and some atmospheric parameters for the 1225 sdO/Bs known at that time. 

Many more hot subdwarfs have been detected subsequently in the Hamburg Quasar Survey (HS, Hagen et al. \cite{hagen95}), the Hamburg ESO survey (HE, Wisotzki et al. \cite{wisotzki96}), the Edinburgh-Cape Survey (EC, Stobie et al. \cite{stobie97}) and the Byurakan surveys (FBS, SBS, Mickaelian et al. \cite{mickaelian07,mickaelian08}). \O stensen (\cite{oestensen06}) did an extensive literature search and created the hot subdwarf database with state-of-art interface linking essentially all the information available in the archives for more than 2300 stars. This database is still widely used in the field.

\begin{figure}[t!]
\begin{center}
	\resizebox{9cm}{!}{\includegraphics{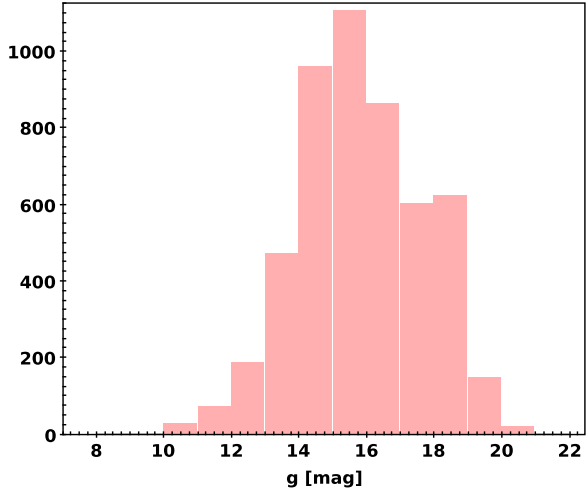}}
\end{center} 
\caption{Magnitude distribution of a representative catalogue subset in the g-band. Photometry has been taken from SDSS if available, otherwise from APASS, which also covers the brighter objects.}
\label{g_dist}
\end{figure}

However, since 2006 the number of known hot subdwarfs again increased by a factor of more than two. The Sloan Digital Sky Survey (SDSS) provided spectra of almost 2000 sdO/Bs (Geier et al. \cite{geier15b}; Kepler et al. \cite{kepler15,kepler16}) reaching down to much fainter magnitudes than previous surveys. On the bright end of the magnitude distribution, new samples of hot subdwarfs have been selected from the EC survey and the GALEX all-sky survey photometry in the UV (e.g. Vennes et al. \cite{vennes11}). Furthermore, new large-area photometric and astrometric surveys have been and are currently conducted in multiple bands from the UV to the far infrared. Given this wealth of new high quality data, we consider it timely to compile a new catalogue of hot subdwarf stars. 

\begin{figure}[t!]
\begin{center}
	\resizebox{9cm}{!}{\includegraphics{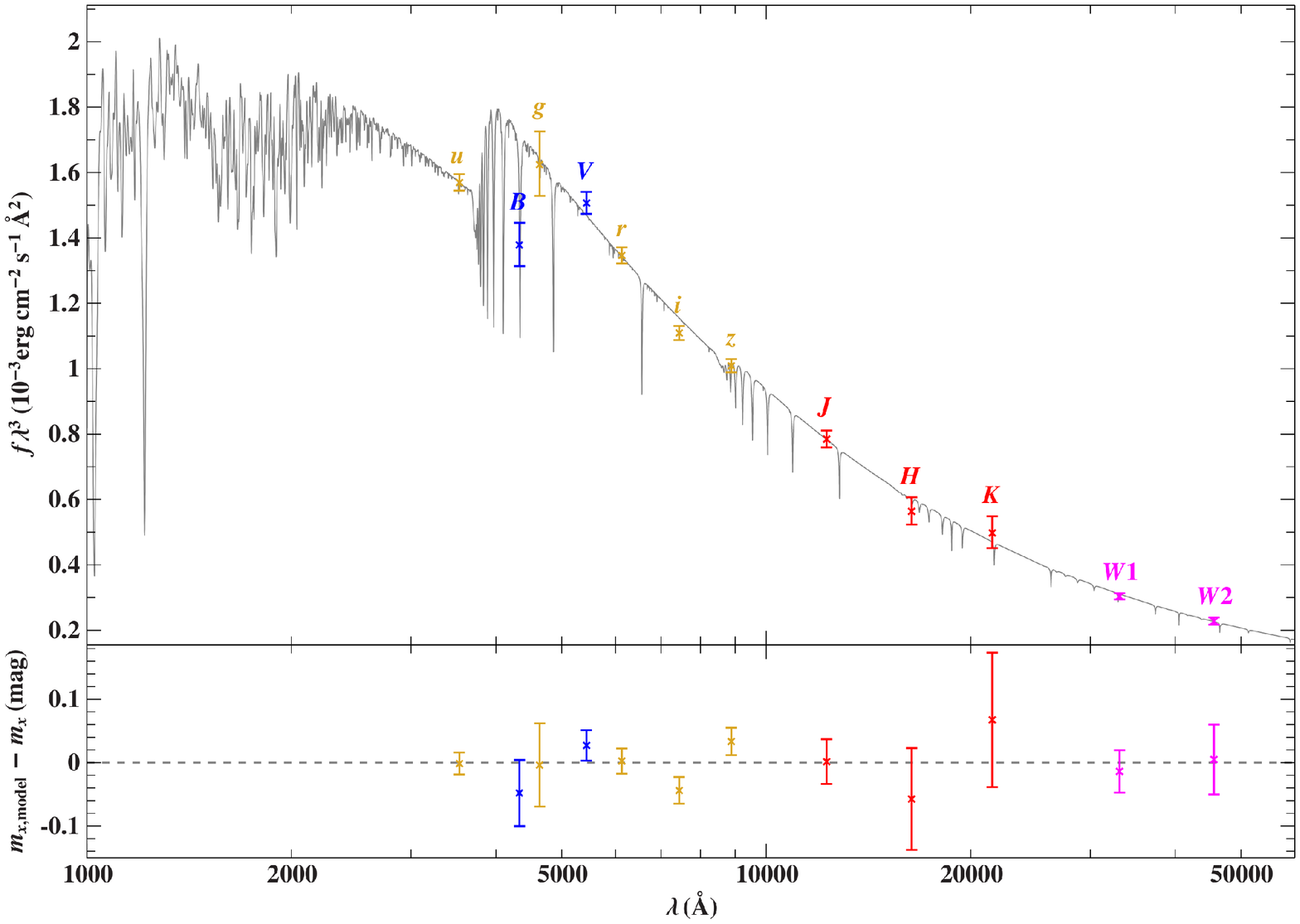}}
	\resizebox{9cm}{!}{\includegraphics{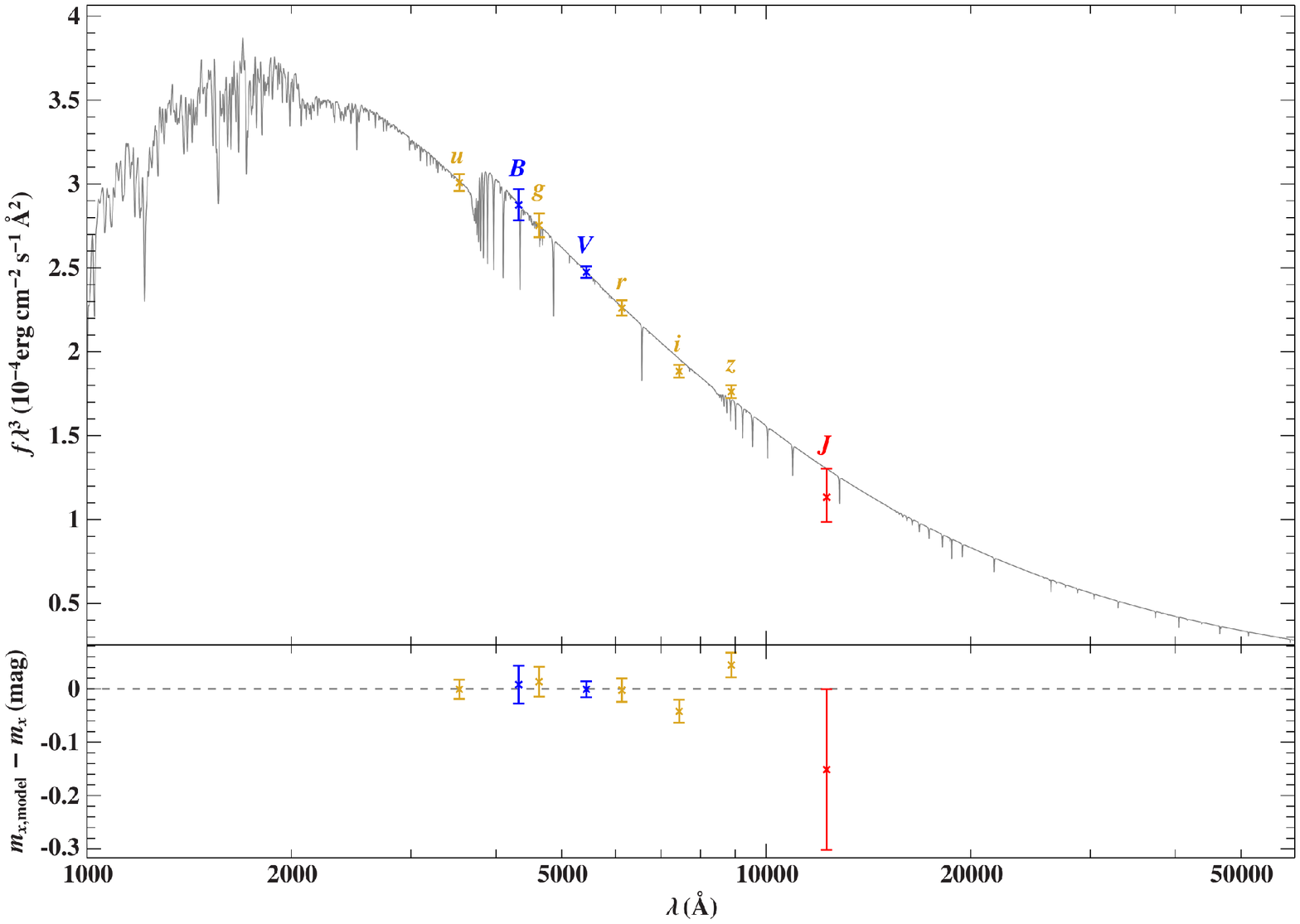}}
	\resizebox{9cm}{!}{\includegraphics{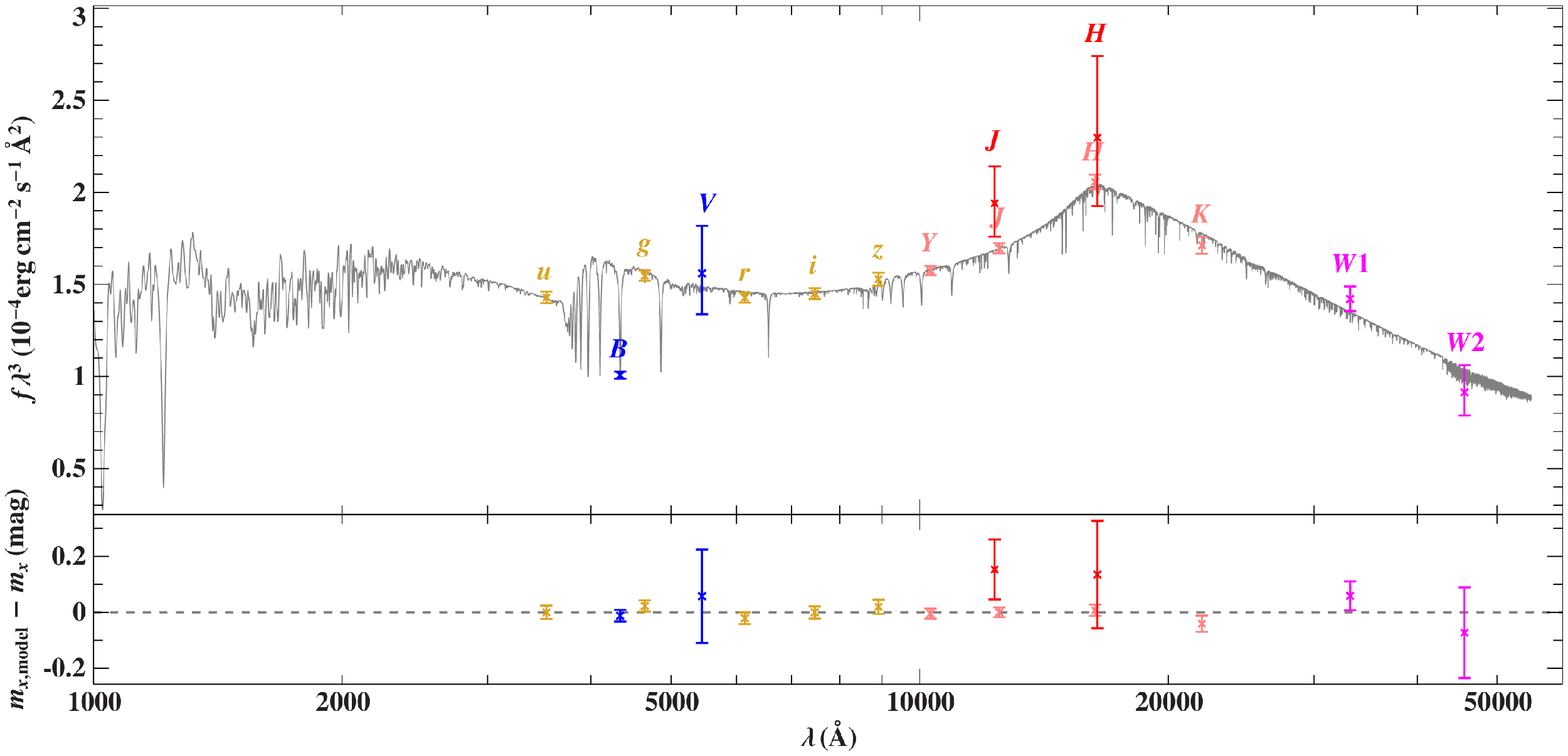}}
\end{center} 
\caption{Preliminary spectral energy distributions and residuals for an sdB star with $T_{\rm eff}\simeq29500\,{\rm K}$ (upper panel), an sdO star with $T_{\rm eff}\simeq39700\,{\rm K}$ (middle panel) and a composite binary system consisting of an sdB star with $T_{\rm eff}\simeq29100\,{\rm K}$ and a K-type main sequence star with $T_{\rm eff}\simeq4500\,{\rm K}$ (lower panel). The photometry is taken from the catalogue. Especially the u-band is crucial to determine reliable effective temperatures for sdO/Bs. The fitting procedure with synthetic models is described in Irrgang (\cite{irrgang14}).}
\label{sed}
\end{figure}

This catalogue will be used as input and calibration dataset to select a magnitude limited, homogeneous, all-sky catalogue of hot subdwarf stars using astrometry and photometry from the Gaia mission, which will allow us to study the properties of the hot subdwarf population with unprecedented accuracy. 

\section{Constructing the catalogue} 

\subsection{Input data}

The basic data source for the catalogue was the sample of hot subdwarfs classified as sdO/B from the database of \O stensen (\cite{oestensen06}), which we consider fairly complete up to the date of publication. We added the subdwarf candidates from the FBS survey (Mickaelian et al. \cite{mickaelian08}), the sample of hot subdwarfs identified in the course of the Kepler mission (\O stensen et al. \cite{oestensen10b}), the large sample of sdO/Bs spectroscopically identified from the Sloan Digital Sky Survey (SDSS) DR7 during the MUCHFUSS project (Massive Unseen Companions to Hot Faint Underluminous Stars from SDSS, Geier et al. \cite{geier15b}), a yet unpublished sample of spectroscopically classified sdO/Bs selected from SDSS DR8-10, and the most recently published sdO/B candidates from SDSS DR12 (Kepler et al. \cite{kepler16}). We also included the candidate sample from SDSS DR10 classified as narrow-line hydrogen stars (NLHS) by Gentile Fusillo et al. (\cite{gentile15}). The recently published large sample of sdO/Bs from the complete EC survey was included as well and is important because of its location in the otherwise underrepresented Southern hemisphere (Stobie et al. \cite{stobie97}; O'Donoghue et al. \cite{odonoghue13}; Kilkenny et al. \cite{kilkenny15,kilkenny16}).  

An important part of our catalogue consists of a sample of several hundred yet unpublished sdO/Bs selected from GALEX, GSC and 2MASS photometry by R.~H. \O stensen and E.~M. Green, which have been classified based on follow-up spectroscopy taken with the INT/IDS, NOT/ALFOSC, WHT/ISIS, CAHA/TWIN, ESO-NTT/EFOSC2, and 4m-KPNO/RC spectrographs. 

In addition, we added the sample of sdO/Bs selected from the Guide Star and the Galaxy Evolution Explorer (GALEX) catalogues by Vennes et al. (\cite{vennes11}), the samples of Oreiro et al. (\cite{oreiro11}) and Perez-Fernandez et al. (\cite{perez16}) selected using Virtual observatory tools and multiband photometry, and the first sample of sdO/Bs discovered by the Large Sky Area Multi-Object Fibre Spectroscopic Telescope (LAMOST) survey (Luo et al. \cite{luo16}). Those samples are considered the most important ones to date and should cover more than $90\%$ of the currently known sdO/Bs in the field. The growing number of hot subdwarf stars found in globular clusters has not been included in this release of the catalogue. 

\subsection{Multi-band photometry}

Using TOPCAT's (Taylor \cite{taylor05}) internal crossmatch with a radius of $10\,{\rm arcsec}$ we identified several hundred duplications and constructed a catalogue of unique sources. To obtain homogeneous multi-band photometry we crossmatched those objects again using a radius of $10\,{\rm arcsec}$ with well calibrated, large-area survey catalogues. NUV and FUV photometry were taken from the GALEX DR5 All-sky Imaging Survey (AIS, Bianchi et al. \cite{bianchi11}). To account for the known systematic shifts of the GALEX magnitudes of bright targets we applied the corrections suggested by Camarota \& Holberg (\cite{camarota14}). 

Optical photometry was obtained from the Guide Star catalogue (GSC~2.3.2, Lasker et al. \cite{lasker08}) in the $R_{\rm F}B_{\rm J}VI_{\rm N}B$-bands, the AAVSO Photometric All Sky Survey (APASS DR9, Henden et al. \cite{henden16}) in the $VBgri$-bands, the SDSS DR12 (Alam et al. \cite{alam15}) in the $ugriz$-bands, the VST-ATLAS (DR2, Shanks et al. \cite{shanks15}) and the Kilo-Degree (KiDS DR2, de Jong et al. \cite{dejong15}) ESO public surveys in the $ugriz$-bands. For the optical magnitude distribution of the catalogue see Fig.~\ref{g_dist}.

\begin{figure}[t!]
\begin{center}
	\resizebox{9cm}{!}{\includegraphics{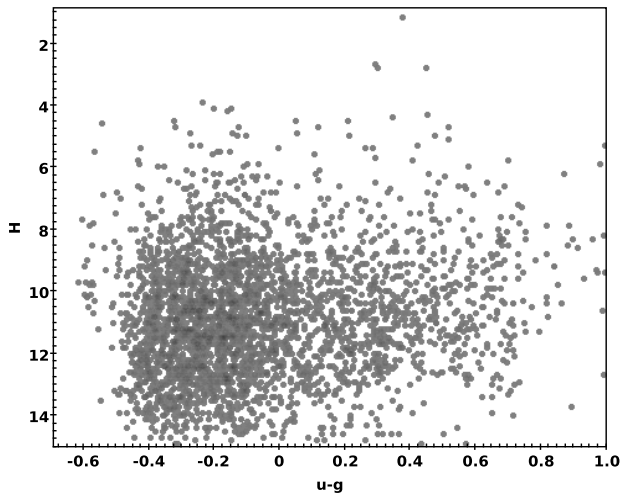}}
\end{center} 
\caption{Reduced proper motion $H$ as defined in Sect.~\ref{cleaning} for a subset of stars with $u-g$ colours. Outliers with inconsistent proper motions between the different proper motion catalogues have been included in our catalogue, but are not shown here.}
\label{red_pm}
\end{figure} 

Near Infrared photometry was obtained from the 2MASS All-Sky Catalog of Point Sources (Skrutskie et al. \cite{skrutskie06}) in the $JHK$-bands, the UKIRT Infrared Deep Sky Survey (UKIDSS Large Area Survey DR9, Lawrence et al. \cite{lawrence07}) in the $YJHK$-bands, the VISTA Hemisphere (VHS DR2, McMahon et al. in prep.), and the VISTA Kilo-degree Infrared Galaxy (VIKING DR4, Edge et al. \cite{edge13}) ESO public surveys in the $ZYJHK_{\rm S}$-bands. Far infrared photometry was obtained from the AllWISE data release (Cutri et al. \cite{cutri14}) in the four WISE-bands.

The Galactic reddening $E(B-V)$ and the Galactic dust extinction $A_V$ from the maps of Schlafly \& Finkbeiner (\cite{schlafly11}) are provided as well. However, correcting for reddening and extinction only works properly for stars situated significantly above the Galactic plane. Bright hot subdwarfs or stars at low Galactic latitudes are usually foreground objects.

The extended wavelength coverage of the multi-band photometry allows us to construct full spectral energy distributions (SEDs) of both single sdO/Bs and sdO/Bs in binaries with cool companions. SEDs are a powerful tool to determine their parameters (e.g. Girven et al. \cite{girven12}). The results of the SED fitting of all the stars in the catalogue will be published in another paper of this series (see Fig.~\ref{sed} for some examples).

\begin{figure*}[t!]
\begin{center}
	\resizebox{17cm}{!}{\includegraphics{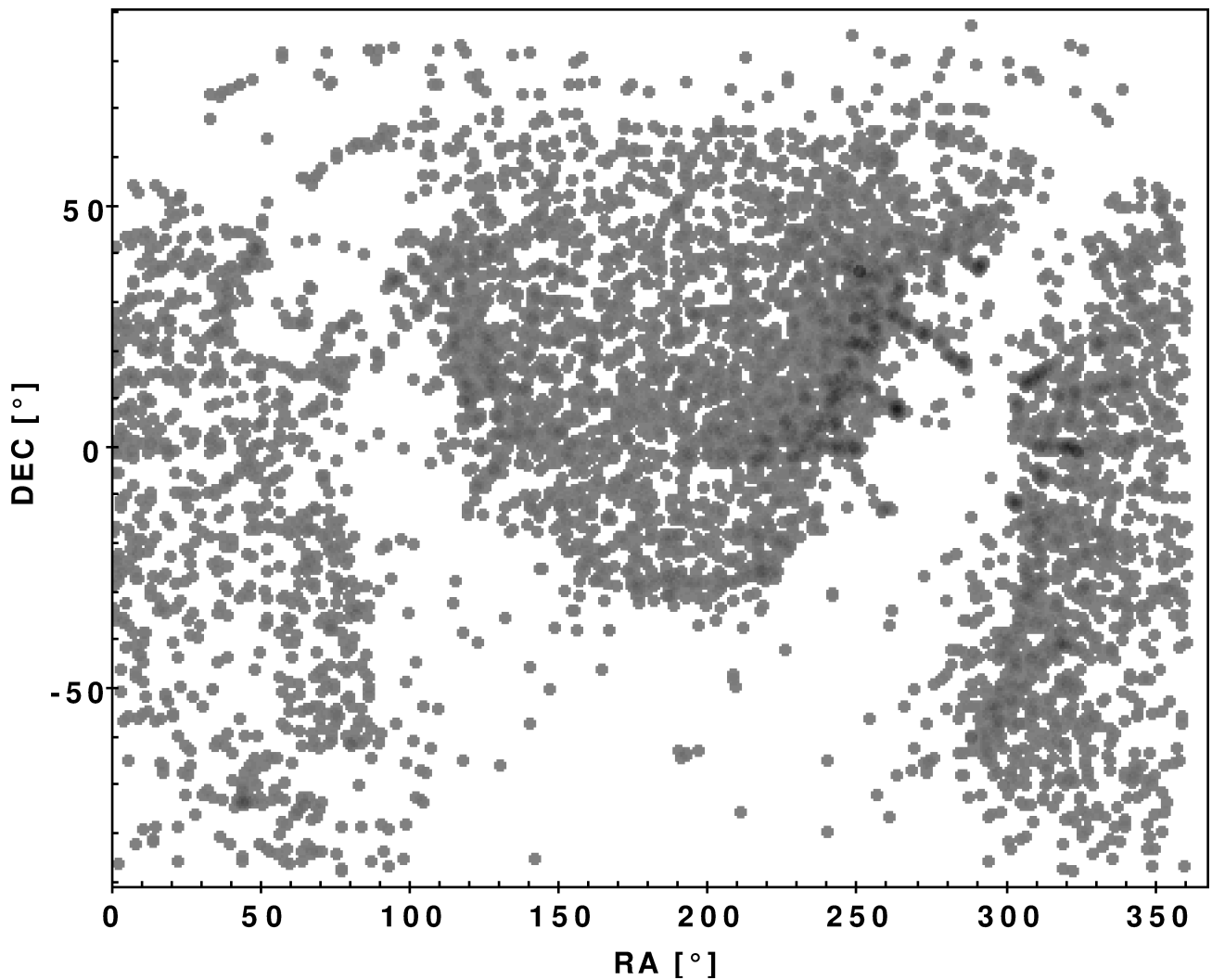}}
\end{center} 
\caption{Coordinates of the objects in the hot subdwarf catalogue.}
\label{footprint}
\end{figure*}

\subsection{Ground-based proper motions}

Ground-based proper motions have been obtained from SDSS (DR9, Ahn et al. \cite{ahn12}), the Fourth U.S. Naval Observatory CCD Astrograph Catalog (UCAC4, Zacharias et al. \cite{zacharias13}), the PPMXL catalog of positions and proper motions on the ICRS (Roeser et al. \cite{roeser10}), the Whole-Sky USNO-B1.0 Catalog (Monet et al. \cite{monet03}), the Absolute Proper motions Outside the Plane catalogue (APOP, Qi et al. \cite{qi15}), and the Yale/San Juan Southern Proper Motion Catalog 4 (SPM4, Girard et al. \cite{girard11}). 

\subsection{Light curve data}

To search for variability likely caused by close binarity (e.g. eclipses, reflection effects, ellipsoidal modulations) or pulsations, available light curves of all the stars in the SuperWASP archive (DR1, Butters et al. \cite{butters10}) and the Catalina Real-Time Transient Survey (CRTS DR2, Drake et al. \cite{drake12}) have been visually inspected. The SuperWASP DR1 only covers a small fraction of the objects in the catalogue and is restricted to bright stars ($<15\,{\rm mag}$). The light curves usually have several thousand single epochs. The CRTS DR2 has a significant overlap with our catalogue and covers the magnitude range $\sim13-20\,{\rm mag}$. The number of epochs varies from a few tens to a few hundred. 

Due to the limited quality of the light curves and strong aliasing effects, we refrained from performing a rigorous statistical analysis of the light curves. Instead the implemented periodogram functions of the Catalina and the NASA Exoplanet Archive webpages were used to search for significant periodicities (keyword {\tt var}). In addition, all light curves were visually checked for irregular variations ({\tt var irregular}). It has to be pointed out that only relatively strong variations can be seen in these data and that objects marked as constant in the catalogue can still show lower amplitude variations. 

The catalogue contains comments about the availability and the properties of the light curves for all stars. Candidates for periodic variabilities come with the tentative period in days (e.g. {\tt var 0.123}). However, only variabilities that are marked with the keyword {\tt strong} should be treated as reliable. The other candidates might well be marginal and require a more sophisticated analysis. We found a variety of new close binary and pulsator candidates, which will be published by Kupfer et al. (in prep.).

\subsection{Cleaning the catalogue}\label{cleaning}

The data collected was used to identify and remove objects misclassified as hot subdwarf stars. Likely candidates are hot white dwarfs of DA, DB and DO type, peculiar objects like PG\,1159 stars and central stars of planetary nebula, but also cooler DAs, which show weaker hydrogen lines and can look like helium-poor sdBs (especially if spectra are normalised to a flat continuum). Main sequence stars of O and B type as well as blue horizontal branch stars are possible bright contaminants. Cool subdwarfs of A and F type as well as cataclysmic variables can mimic sdO/B+MS binaries. Extragalactic objects like blue galaxies and QSOs also appear in colour-selected samples. To separate all kinds of cool objects, colour indices were used. About $300$ objects with SDSS colours $u-g>0.6$ and $g-r>0.1$ as well as $NUV_{\rm GALEX}-g_{\rm APASS}>2.0$ have been excluded. 

A powerful tool to separate nearby white dwarfs from the more luminous and distant sdO/Bs is the reduced proper motion. We followed the approach outlined in Gentile Fusillo et al. (\cite{gentile15}) and calculated the reduced proper motion $H=x+5\log{\mu }+5$. The full proper motion $\mu$ was averaged from all available proper motions of each star. Since not all stars have photometry in all bands, we defined the magnitude $x$ as the average of all the blue and visual bands in the catalogue ($Bj_{\rm GSC}$, $V_{\rm GSC}$, $B_{\rm GSC}$, $V_{\rm APASS}$, $B_{\rm APASS}$, $g_{\rm APASS}$, $u_{\rm SDSS}$, $g_{\rm SDSS}$, $u_{\rm VST}$, $g_{\rm VST}$). It has to be pointed out that this is not a physically meaningful quantity, but an empirically determined filter parameter only. For the subset of stars with SDSS photometry, we compared $H$ with the reduced proper motion in the g-band $H_{\rm g}=g+5\log{\mu }+5$. The mean deviation is $\pm0.12$, which can become much higher ($>0.5$) for stars with incomplete photometry.

We constructed the reduced proper motion diagram and also included previously missclassified white dwarfs (WDs) to find the most reasonable exclusion criterion (see Fig.~\ref{red_pm}). We found that stars with $H>15$ are very likely to be WDs (see also Gentile Fusillo et al. \cite{gentile15}) and excluded them. Exceptions have been made for stars with highly inconsistent proper motions.  

\begin{figure}[t!]
\begin{center}
	\resizebox{9.5cm}{!}{\includegraphics{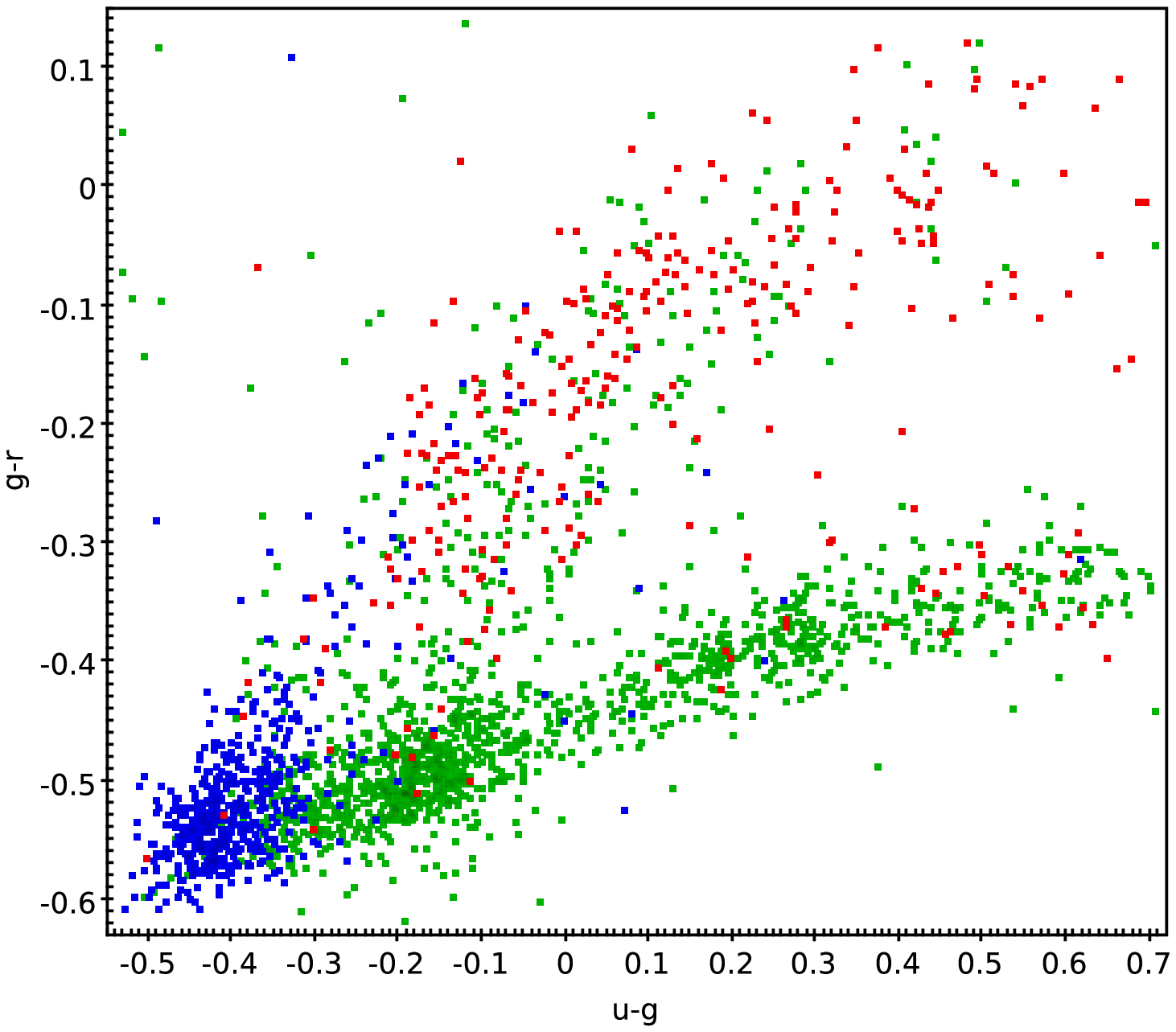}}
	\resizebox{9.5cm}{!}{\includegraphics{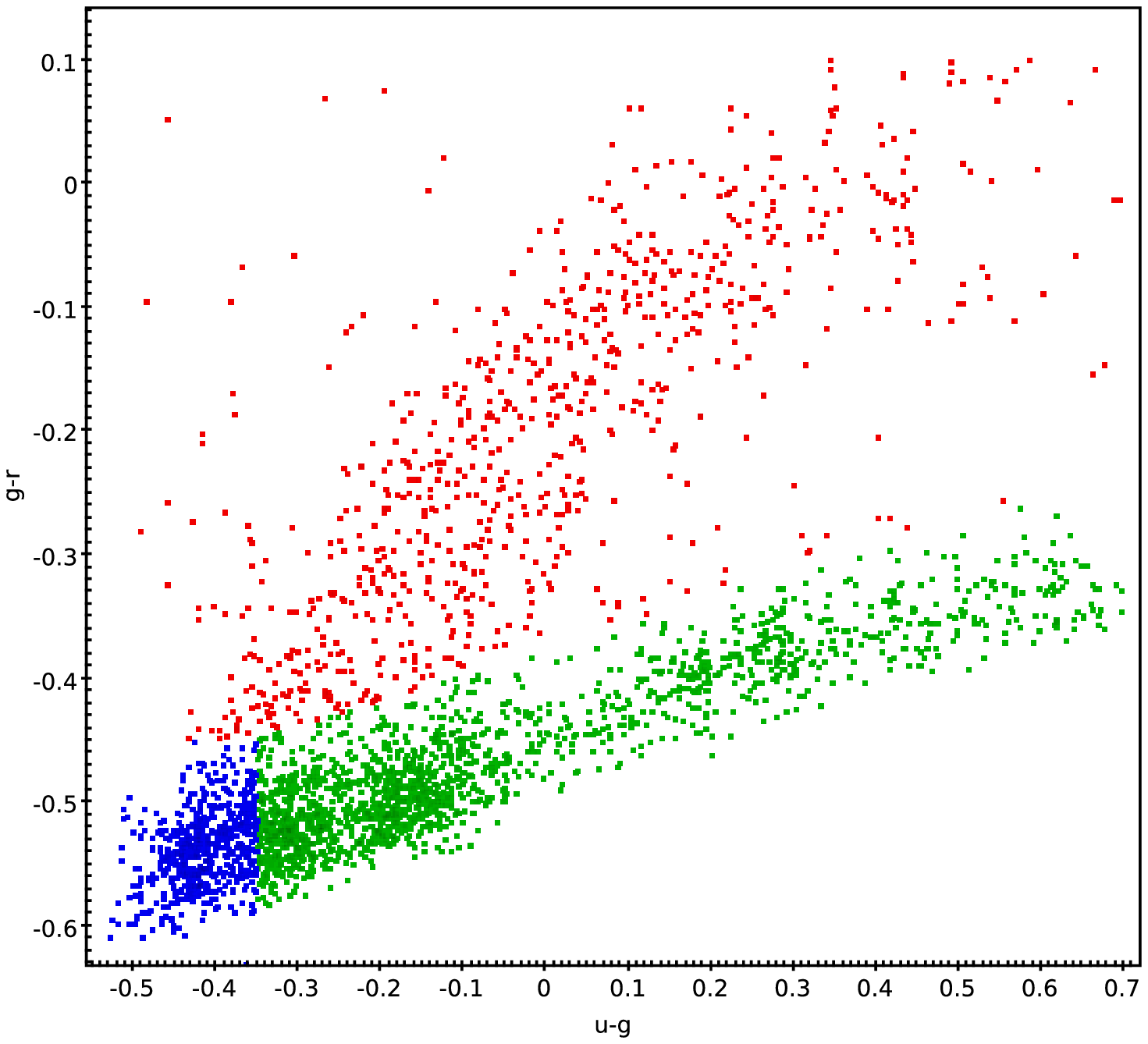}}
\end{center} 
\caption{Two colour diagram for the stars with SDSS colour classifications. The colour indices have been corrected for reddening as described in Schlafly \& Finkbeiner (\cite{schlafly11}). Stars classified in this way as sdOs are marked in blue, sdBs in green and composite sd+MS systems in red. The upper panel shows stars with spectroscopic classifications. The colour cuts in the lower panel are determined from this plot. It can be clearly seen that composite systems are easier to identify from their colours than from their spectra.}
\label{class1}
\end{figure} 

\begin{figure}[t!]
\begin{center}
	\resizebox{9cm}{!}{\includegraphics{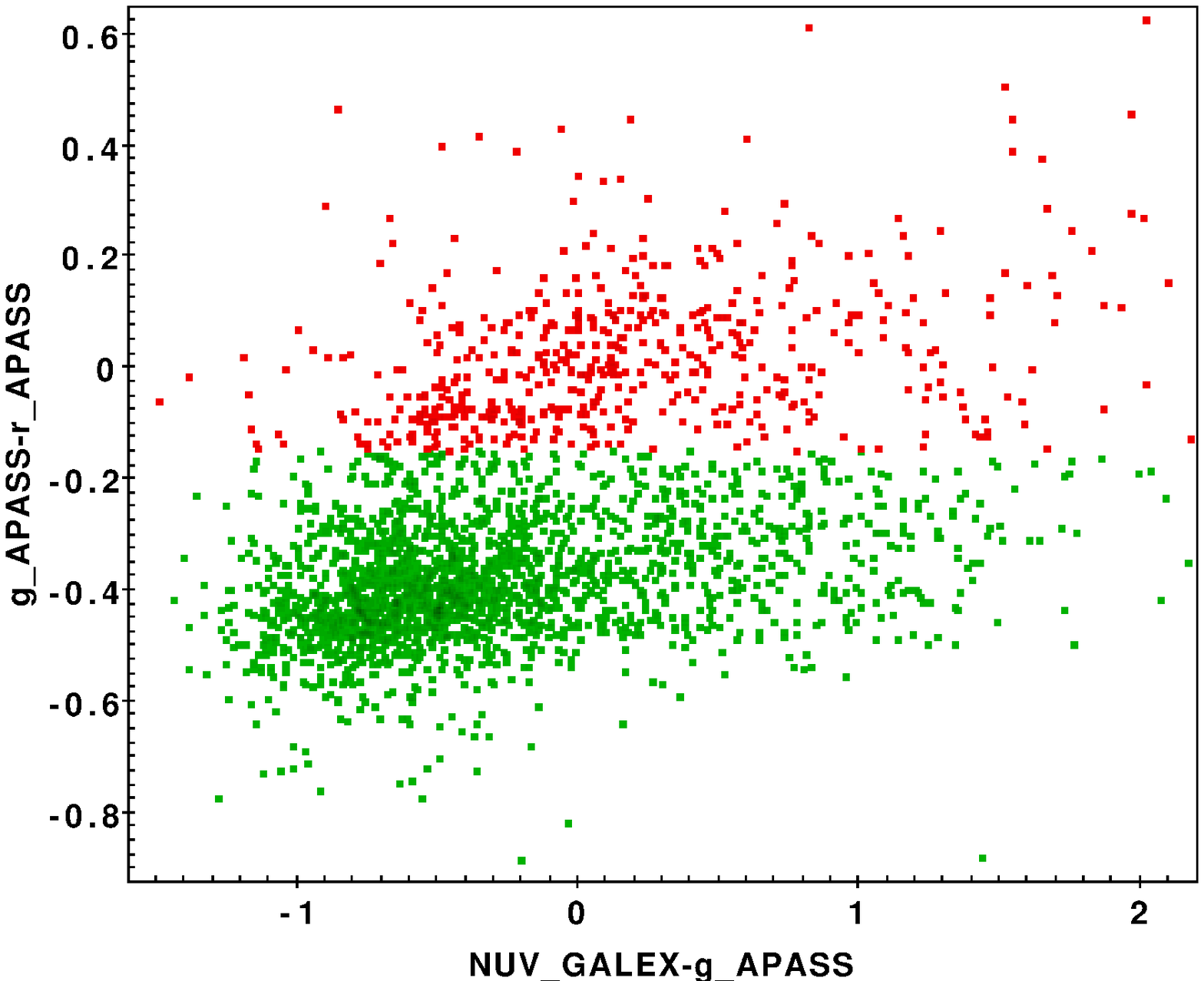}}
        \resizebox{9cm}{!}{\includegraphics{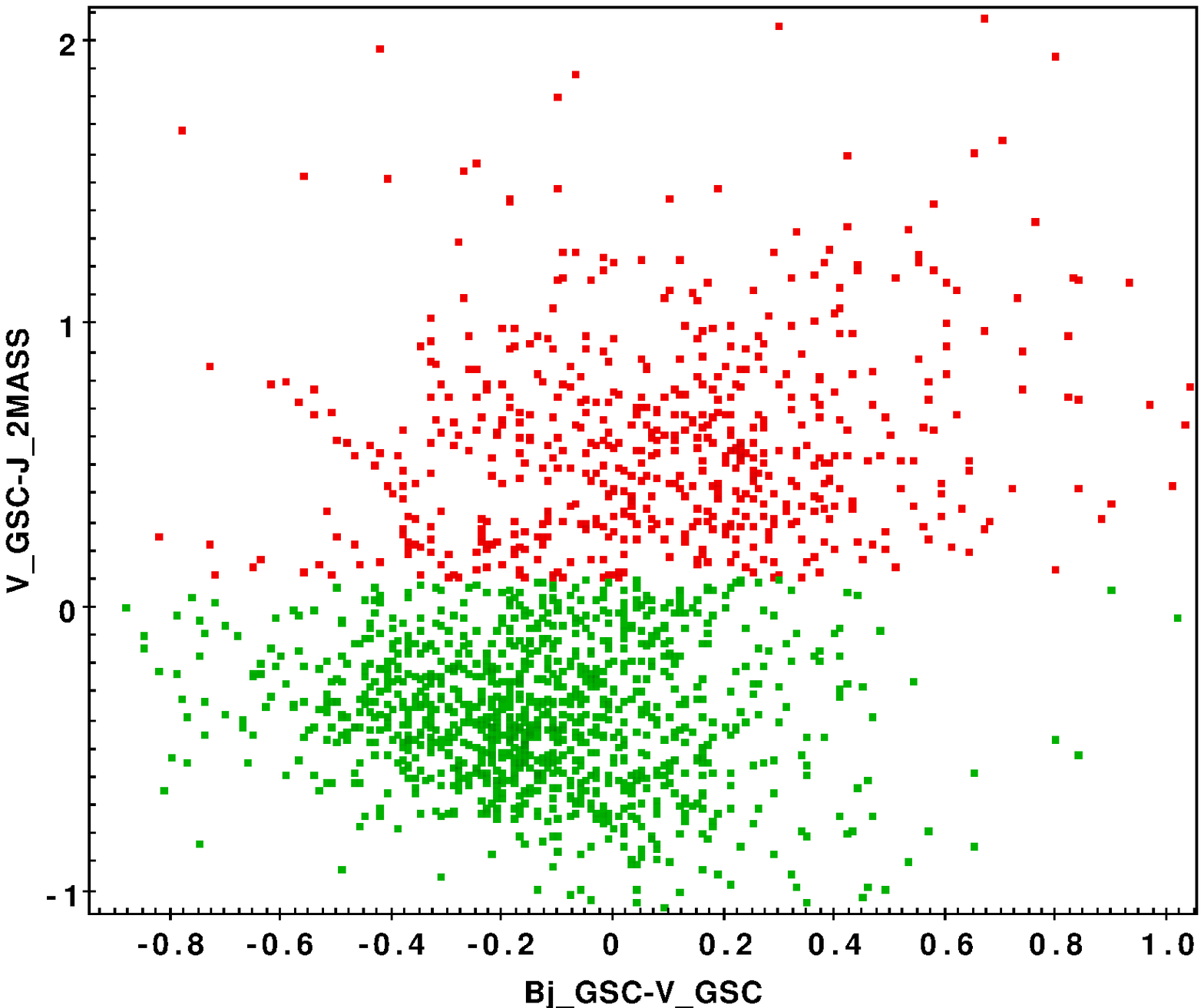}}
\end{center} 
\caption{Two colour diagrams for the stars with GALEX/APASS (upper panel) and GSC/2MASS (lower panel) colour classifications. Because those stars are on average brighter and closer, no reddening correction has been applied. Stars classified in this way as sdO/Bs are marked in green and composite sd+MS systems in red.}
\label{class2}
\end{figure} 

The light curves also turned out to be useful to identify misclassifications. Strong irregular variations led to the exclusion of some cataclysmic variables (CVs) and the characteristic pulsations of RR\,Lyr variables are easy to recognize and allowed us to exclude a few tens of those stars. 

The catalogue was also crossmatched with SIMBAD and misclassified objects known from the literature (mostly B-stars, WDs, CVs, and central stars of planetary nebula) were excluded as well. Another benefit of the various crossmatching exercises was the identification of a few tens of stars (some of them very bright) with wrong coordinates in SIMBAD. We corrected those coordinates in the catalogue by adopting either SDSS or 2MASS positions. 

Finally, we inspected the distribution of values in the single catalogue columns as well as the two colour diagrams with TOPCAT to find and eliminate obvious outliers. However, the large amount of heterogeneous archive data has not been systematically checked for quality (e.g. by inspecting quality flags). The final catalogue contains 5613 unique objects, most of which are located in the northern hemisphere and at rather high Galactic latitudes, because most surveys cover those regions (see Fig.~\ref{footprint}). Besides some WDs and MS-B stars we expect blue horizontal branch (BHB) stars to be the most important class of contaminant objects remaining in the catalogue, because they are very hard to separate from sdO/Bs with the methods used here.

First preliminary data from the Gaia mission has been published recently (Gaia collaboration \cite{gaia16}). However, due to yet uncorrected chromatic effects, the bluest objects have been excluded from Gaia Data Release 1. Only $\sim60$ stars of our catalogue are part of the bright TGAS sample (Tycho-Gaia astrometric solution, Michalik et al. \cite{michalik15}) with preliminary parallaxes and proper motions, almost all of them composite sdB binaries. The much larger Gaia DR1 catalogue of positions and Gaia magnitudes only covers about half of our sample. Due to this severe selection bias we decided not to include Gaia DR1 data in this version of our catalogue. 

\subsection{Classification of hot subdwarfs}

Most of the stars in the catalogue have been visually classified by looking at some kind of optical spectrum (objective prism, long-slit, echelle). It is almost impossible to homogenize the classifications attributed by different people at different times based on very different types of data. Furthermore, quite diverse classification schemes emerged in the last decades. From the quite detailed classifications in the PG catalogue (Green et al. \cite{green86}) to the more general classes introduced (and widely used today) by Moehler et al. (\cite{moehler90}) to the sophisticated MK-like classification system developed by Drilling et al. (\cite{drilling13}).

The spectroscopic classifications provided for 5055 stars in the catalogue (see Table~\ref{tab1}) follow the scheme outlined in Moehler et al. (\cite{moehler90}). Classifications from the literature have been adapted to this scheme. The reason is that statistical ana\-lyses of population properties require a meaningful definition of the sub-populations rather than too detailed classifications. The exact spectroscopic class might still be interesting to study single objects. In this case, the catalogue links to the literature can be used to find out more.

It has to be pointed out that some subclasses are rather difficult to distinguish spectroscopically. The only difference between the sdBs and sdOBs is often the presence or absence of only one He\,{\sc ii} line at $4686\,{\rm \AA}$, which can be easily overlooked in low quality spectra. The helium-rich subclasses (He-sdB, especially He-sdOB and He-sdO) can often only be distinguished with a proper quantitative spectral analysis. Stars classified as sdO in the catalogue can be both hydrogen- or helium-rich, because they are often not clearly separated in the literature.

Assuming that the spectroscopic classifications are on average correct, we used them to define a purely empirical scheme for the photometric classifications by inspecting the locations of the subclasses in two-colour diagrams (see Table~\ref{tab2}). Using the SDSS $u-g$ versus $g-r$ diagram (Fig.~\ref{class1}, see also Geier et al. \cite{geier11a}), sdOs of all types are well separated from the sdBs. sd+MS stars on the other hand form a distinct sequence and are clearly separated from the cooler stars of A and F-type, which have been excluded. The sd+MS sequence is also well defined in the $NUV_{\rm GALEX}-g_{\rm APASS}$ versus $g_{\rm APASS}-r_{\rm APASS}$ diagram (Fig.~\ref{class2}, upper panel), whereas the hotter sdOs cannot be separated. To provide at least a rough photometric class of the remaining objects, we also used the less well defined colour criterium $V_{\rm GSC}-J_{\rm 2MASS}$ (Fig.~\ref{class2}, lower panel) to distinguish between single sdO/Bs and composite sd+MS systems. The SDSS based colour class should be regarded as most trustworthy, followed by the GALEX/APASS class and eventually the GSC/2MASS class. Following this ranking, we defined a comprehensive colour class and classified 4888 stars in the catalogue in this way.

The colour classes can also used as a sanity check for the spectroscopic classifications, e.g. to find yet undetected main sequence companions, and of course are very useful for a tentative   classification of stars without spectroscopy.

\begin{table}
\caption{\label{tab1} Spectral classification scheme\tablefootmark{a}}
\begin{center}
\begin{tabular}{ll}
\hline\hline
\noalign{\smallskip}
sdB & H, (He\,{\sc i}) \\
sdOB & H, He\,{\sc i}, He\,{\sc ii} \\
sdO & H, He\,{\sc ii} (H-rich) or He\,{\sc ii}, (H) (He-rich) \\
He-sdB & He\,{\sc i}, (H) \\
He-sdOB & He\,{\sc i}, He\,{\sc ii}, (H) \\
He-sdO & He\,{\sc ii}, (He\,{\sc i}, H) \\
sdO/B+K/G/F/A & sdO/B with visible MS companion \\
\noalign{\smallskip}
\hline
\noalign{\smallskip}
sdO/BV & Pulsator \\
sdO/B+WD/dM/BD & sdO/B with close invisible companion \\
\noalign{\smallskip}
\hline\hline
\end{tabular}
\end{center}
\tablefoot{
\tablefoottext{a}{For each class the species with observed lines in the spectrum are given in a sequence of relative strength. The strongest lines are always coming first. 
Species in parentheses can be present, but do not have to be.}
}
\end{table} 

\begin{table}
\caption{\label{tab2} Colour classification schemes}
\begin{center}
\begin{tabular}{ll}
\hline\hline
\noalign{\smallskip}
Colour class 1 & \\
SDSS & \\
\noalign{\smallskip}
\hline
\noalign{\smallskip}
sdO & $-0.55<u_{\rm SDSS}-g_{\rm SDSS}<-0.35$ \\
    & $-0.65<g_{\rm SDSS}-r_{\rm SDSS}<-0.45$ \\
sdB & $-0.5<u_{\rm SDSS}-g_{\rm SDSS}<0.7$ \\
    & $g_{\rm SDSS}-r_{\rm SDSS}>0.208(u_{\rm SDSS}-g_{\rm SDSS})-0.516$ \\
    & $g_{\rm SDSS}-r_{\rm SDSS}<0.208(u_{\rm SDSS}-g_{\rm SDSS})-0.376$ \\
sd+MS & $-0.5<u_{\rm SDSS}-g_{\rm SDSS}<0.7$ \\
      & $g_{\rm SDSS}-r_{\rm SDSS}>0.208(u_{\rm SDSS}-g_{\rm SDSS})-0.376$ \\
\noalign{\smallskip}
\hline
\noalign{\smallskip}
Colour class 2 & \\
GALEX/APASS & \\
\noalign{\smallskip}
\hline
\noalign{\smallskip}
sdO/B & $NUV_{\rm GALEX}-g_{\rm APASS}<2.0$ \\
      & $g_{\rm APASS}-r_{\rm APASS}<-0.15$ \\
sd+MS & $NUV_{\rm GALEX}-g_{\rm APASS}<2.0$ \\
      & $g_{\rm APASS}-r_{\rm APASS}\geq-0.15$ \\
\noalign{\smallskip}
\hline
\noalign{\smallskip}
Colour class 3 & \\
GSC/2MASS & \\
\noalign{\smallskip}
\hline
\noalign{\smallskip}
sdO/B & $V_{\rm GSC}-J_{\rm 2MASS}<0.1$ \\
sd+MS & $V_{\rm GSC}-J_{\rm 2MASS}\geq0.1$ \\
\noalign{\smallskip}
\hline\hline
\end{tabular}
\end{center}
\end{table} 

\subsection{Spectroscopic parameters and radial velocities}

The catalogue contains spectroscopic parameters like effective temperatures, surface gravities and helium abundances for 924 stars from the literature. This fraction is not complete, because only papers containing larger samples of sdO/B stars have been taken into account (Heber et al. \cite{heber84}; Bixler et al. \cite{bixler91}; Saffer et al. \cite{saffer94,saffer97}; Maxted et al. \cite{maxted01}; Edelmann et al. \cite{edelmann03}; Lisker et al. \cite{lisker05}; Str\"oer et al. \cite{stroeer07}; Hirsch \cite{hirsch09}; \O stensen et al. \cite{oestensen10a}; Nemeth et al. \cite{nemeth12}; Geier et al. \cite{geier13};  Geier et al. \cite{geier15b}; Kupfer et al. \cite{kupfer15}; Luo et al. \cite{luo16}; Kepler et al. \cite{kepler16}). 

Radial velocities (RVs) are provided for the 2122 stars with spectra in the SDSS data archive. Those RVs have been measured automatically by crosscorrelation with template spectra in the course of the SDSS. Since the template library does not include very helium-rich stars like He-sdOs, the RVs of such stars can be wrong by up to $\sim100\,{\rm km\,s^{-1}}$. The reason for this is that the Pickering series of single ionized helium can easily be confused with the hydrogen Balmer series. However, the rest wavelengths of the Pickering lines are blueshifted with respect to the Balmer lines. Fitting a He-sdO with a normal O star template can therefore lead to an apparently decent match, but result in the wrong RV. For He-sdOs with pure helium atmospheres this effect is strongest and must be taken into account (see Geier et al. \cite{geier15a}).

The RVs are obtained from SDSS spectra, which are usually coadded from three consecutive individual exposures of $15\,{\rm min}$ duration. Since especially sdB stars are often found in close binaries with sometimes very short orbital periods, orbital smearing can lead to systematic shifts (e.g. Geier et al. \cite{geier11a}). More importantly, the SDSS RVs in this catalogue are only obtained from one coadded spectrum and contain no information about RV variability (as provided for sdO/Bs up to SDSS DR7 in Geier et al. \cite{geier15b}). 

\begin{table}
\caption{\label{tab3} Catalogue statistics}
\begin{center}
\begin{tabular}{ll}
\hline\hline
\noalign{\smallskip}
Total & 5613 \\
\noalign{\smallskip}
\hline
\noalign{\smallskip}
Spectroscopic class & \\
\noalign{\smallskip}
\hline
\noalign{\smallskip}
sdB & 2866 \\
sdOB & 530 \\
sdO & 477 \\
He-sdB/OB/O & 540 \\
sd+MS & 642 \\
\noalign{\smallskip}
\hline
\noalign{\smallskip}
Unclassified sd & 558 \\
\noalign{\smallskip}
\hline
\noalign{\smallskip}
Colour class & \\
\noalign{\smallskip}
\hline
\noalign{\smallskip}
sdB & 1666 \\
sdO & 548 \\
sdO/B & 1448 \\
sd+MS & 1226 \\
\noalign{\smallskip}
\hline
\noalign{\smallskip}
Unclassified sd & 725 \\
\noalign{\smallskip}
\hline\hline
\end{tabular}
\end{center}
\end{table} 

\subsection{Catalogue statistics and data access}

Table~3 summarizes the content of the catalogue. Since the sample is heterogeneously selected, those numbers only contain very limited information about the quantitative properties of the underlying Galactic hot subdwarf population. The most striking difference between the spectroscopic and the colour selection is the much higher number of composite sd+MS systems found by colours. Part of this mismatch can very likely be attributed to the quality of the photometry and the selection of the colour cuts. 

However, considering that most spectral classifications are based on blue and visual spectra, where the cool companion is quite often outshone by the sdO/B component, infrared excesses are a much better indication for a cool companion than the sometimes quite shallow spectral lines (e.g. Nemeth et al. \cite{nemeth16}), we conclude that the colour selection should be better suited to find such binaries especially if the companions are of late type and that the fraction of those systems might have been underestimated so far.  

The catalogue will be available via the VizieR service as well as the German Virtual Observatory (GAVO). A detailed description of the catalogue columns is provided in Table~\ref{tab3}. It will be linked to all relevant databases to allow an easy access to data for individual objects. The catalogue is by no means complete and heterogeneously selected, which has to be taken into account, when using it for statistical analyses. People who are interested in individual objects should carefully check the relevant quality flags in the specific catalogues. 

If an object is in the catalogue, there is a high chance that it will indeed be a hot subdwarf star. However, some misclassified objects are certainly still there. If a hot subdwarf candidate from one of the input samples should not be in the catalogue, it is extremely likely, that it is not a hot subdwarf star, because it did not pass our selection criteria. Combining the colour classes with the spectroscopic classification it is possible to select purer subsamples. 

\section{Conclusions}

The new catalogue of hot subdwarfs presented here contains a significant fraction of the total sdO/B population with apparent visual magnitudes between $\sim9\,{\rm mag}$ and $\sim20\,{\rm mag}$, which translates into distances from a few ten pc to more than $20\,{\rm kpc}$ and therefore includes stars from the thin disk, thick disk and halo population. This reasonably well characterised sample will be crossmatched with the Gaia catalogue and used to define the criteria (reduced proper motions, distances, colour cuts in the Gaia bands, etc.) for the selection of a homogeneous all-sky catalogue of sdO/B stars. 

It will become an important input catalogue for ground-based light curve transient and transit surveys like the Palomar Transient Factory (PTF), the BlackGEM and Gravitational-wave Optical Transient Observer (GoTo) surveys for optical counterparts of gravitational wave transients, the Next Generation Transit Survey for exoplanets (NGTS), or the Large Synoptic Survey Telescope (LSST), but also space missions like the ongoing K2 mission, the PLAnetary Transits and Oscillations of stars (PLATO) mission, the Transiting Exoplanet Survey Satellite (TESS), the Wide Field Infrared Survey Telescope (WFIRST), and the Euclid mission. Due to the high fraction of close binaries in the hot subdwarf population we expect to find many of those based on their characteristic light curve variations. 

We are planning to use the catalogue also as input for wide-area spectroscopic surveys like LAMOST, the WEAVE survey at La Palma, the 4-metre Multi-Object Spectroscopic Telescope (4MOST) survey in Chile, and Dark Energy Spectroscopic Instrument (DESI) survey and obtain spectroscopy for a large fraction of the stars. With a density of less than 1 object per square degree on average, only very few fibres would be necessary to carry out a fairly complete survey as side project to the main surveys.

Besides the extension of the catalogue and the inclusion of the Gaia data, other important datasets like the multi-band, wide-area surveys PanSTARRS and SkyMapper will be included in future releases. 

\begin{acknowledgements}

We want to thank the referee Dave Kilkenny for his constructive report and for reminding us not to neglect the southern hemisphere.

The research leading to these results has received funding from the European Research Council under the European Union’s Seventh Framework Programme (FP/2007–2013) / ERC Grant Agreement n. 320964 (WDTracer).

This research made use of TOPCAT, an interactive graphical viewer and editor for tabular data Taylor (\cite{taylor05}). This research made use of the SIMBAD database, operated at CDS, Strasbourg, France; the VizieR catalogue access tool, CDS, Strasbourg, France. Some/all of the data presented in this paper were obtained from the Mikulski Archive for Space Telescopes (MAST). STScI is operated by the Association of Universities for Research in Astronomy, Inc., under NASA contract NAS5-26555. Support for MAST for non-HST data is provided by the NASA Office of Space Science via grant NNX13AC07G and by other grants and contracts. This research has made use of the services of the ESO Science Archive Facility.

This publication makes use of data products from the Two Micron All Sky Survey, which is a joint project of the University of Massachusetts and the Infrared Processing and Analysis Center/California Institute of Technology, funded by the National Aeronautics and Space Administration and the National Science Foundation. Based on observations made with the NASA Galaxy Evolution Explorer. GALEX is operated for NASA by the California Institute of Technology under NASA contract NAS5-98034. This research has made use of the APASS database, located at the AAVSO web site. Funding for APASS has been provided by the Robert Martin Ayers Sciences Fund. The Guide Star Catalogue-II is a joint project of the Space Telescope Science Institute and the Osservatorio Astronomico di Torino. Space Telescope Science Institute is operated by the Association of Universities for Research in Astronomy, for the National Aeronautics and Space Administration under contract NAS5-26555. The participation of the Osservatorio Astronomico di Torino is supported by the Italian Council for Research in Astronomy. Additional support is provided by European Southern Observatory, Space Telescope European Coordinating Facility, the International GEMINI project and the European Space Agency Astrophysics Division.

Based on observations obtained as part of the VISTA Hemisphere Survey, ESO Program, 179.A-2010 (PI: McMahon). This publication has made use of data from the VIKING survey from VISTA at the ESO Paranal Observatory, programme ID 179.A-2004. Data processing has been contributed by the VISTA Data Flow System at CASU, Cambridge and WFAU, Edinburgh. Based on data products from observations made with ESO Telescopes at the La Silla Paranal Observatory under program ID 177.A 3011(A,B,C,D,E.F). Based on data products from observations made with ESO Telescopes at the La Silla Paranal Observatory under programme IDs 177.A-3016, 177.A-3017 and 177.A-3018, and on data products produced by Target/OmegaCEN, INAF-OACN, INAF-OAPD and the KiDS production team, on behalf of the KiDS consortium. OmegaCEN and the KiDS production team acknowledge support by NOVA and NWO-M grants. Members of INAF-OAPD and INAF-OACN also acknowledge the support from the Department of Physics \& Astronomy of the University of Padova, and of the Department of Physics of Univ. Federico II (Naples). This publication makes use of data products from the Wide-field Infrared Survey Explorer, which is a joint project of the University of California, Los Angeles, and the Jet Propulsion Laboratory/California Institute of Technology, and NEOWISE, which is a project of the Jet Propulsion Laboratory/California Institute of Technology. WISE and NEOWISE are funded by the National Aeronautics and Space Administration.

Based on observations made with the Nordic Optical Telescope, operated by the Nordic Optical Telescope Scientific Association at the Observatorio del Roque de los Muchachos, La Palma, Spain, of the Instituto de Astrofisica de Canarias. Based on observations at the La Silla-Paranal Observatory of the European Southern Observatory. Based on observations collected at the Centro Astron\'omico Hispano Alem\'an (CAHA) at Calar Alto, operated jointly by the Max-Planck Institut f\"ur Astronomie and the Instituto de Astrof\'isica de Andaluc\'ia (CSIC). Based on observations with the William Herschel and Isaac Newton Telescopes operated by the Isaac Newton Group at the Observatorio del Roque de los Muchachos of the Instituto de Astrofisica de Canarias on the island of La Palma, Spain. Based on observations at Kitt Peak National Observatory, National Optical Astronomy Observatory, which is operated by the Association of Universities for Research in Astronomy (AURA) under cooperative agreement with the National Science Foundation. The authors are honored to be permitted to conduct astronomical research on Iolkam Du'ag (Kitt Peak), a mountain with particular significance to the Tohono O'odham. 

Funding for the SDSS and SDSS-II has been provided by the Alfred P. Sloan Foundation, the Participating Institutions, the National Science Foundation, the U.S. Department of Energy, the National Aeronautics and Space Administration, the Japanese Monbukagakusho, the Max Planck Society, and the Higher Education Funding Council for England. The SDSS Web Site is http://www.sdss.org/. The SDSS is managed by the Astrophysical Research Consortium for the Participating Institutions. The Participating Institutions are the American Museum of Natural History, Astrophysical Institute Potsdam, University of Basel, University of Cambridge, Case Western Reserve University, University of Chicago, Drexel University, Fermilab, the Institute for Advanced Study, the Japan Participation Group, Johns Hopkins University, the Joint Institute for Nuclear Astrophysics, the Kavli Institute for Particle Astrophysics and Cosmology, the Korean Scientist Group, the Chinese Academy of Sciences (LAMOST), Los Alamos National Laboratory, the Max-Planck-Institute for Astronomy (MPIA), the Max-Planck-Institute for Astrophysics (MPA), New Mexico State University, Ohio State University, University of Pittsburgh, University of Portsmouth, Princeton University, the United States Naval Observatory, and the University of Washington. 

Funding for SDSS-III has been provided by the Alfred P. Sloan Foundation, the Participating Institutions, the National Science Foundation, and the U.S. Department of Energy Office of Science. The SDSS-III web site is http://www.sdss3.org/. SDSS-III is managed by the Astrophysical Research Consortium for the Participating Institutions of the SDSS-III Collaboration including the University of Arizona, the Brazilian Participation Group, Brookhaven National Laboratory, University of Cambridge, Carnegie Mellon University, University of Florida, the French Participation Group, the German Participation Group, Harvard University, the Instituto de Astrofisica de Canarias, the Michigan State/Notre Dame/JINA Participation Group, Johns Hopkins University, Lawrence Berkeley National Laboratory, Max Planck Institute for Astrophysics, Max Planck Institute for Extraterrestrial Physics, New Mexico State University, New York University, Ohio State University, Pennsylvania State University, University of Portsmouth, Princeton University, the Spanish Participation Group, University of Tokyo, University of Utah, Vanderbilt University, University of Virginia, University of Washington, and Yale University. 

\end{acknowledgements}

\longtab{4}{
\begin{longtable}{llll}
\caption{\label{tab3} Catalog columns}\\
\hline\hline
\noalign{\smallskip}
Column & Format & Description & Unit \\
\noalign{\smallskip}
\hline
\noalign{\smallskip}
NAME & A30 & Target name & \\
RA & F10.6 & Right ascension (J2000) & deg \\
DEC & F10.6 & Declination (J2000) & deg \\
SPEC\_CLASS & A15 & Spectroscopic classification & \\
COLOUR\_CLASS1 & A10 & Colour classification SDSS & \\
COLOUR\_CLASS2 & A10 & Colour classification GALEX/APASS & \\
COLOUR\_CLASS3 & A10 & Colour classification GSC/2MASS & \\
COLOUR\_CLASS & A10 & Colour classification  & \\
RV\_SDSS & F5.1 & Radial velocity SDSS & ${\rm km\,s^{-1}}$ \\
e\_RV\_SDSS & F5.1 & Error on RV\_SDSS & ${\rm km\,s^{-1}}$ \\
T\_EFF & F8.1 & Effective temperature & K \\
e\_T\_EFF & F8.1 & Error on T\_EFF & K \\
LOG\_G & F4.2 & Log surface gravity (gravity in ${\rm cm\,s^{-2}}$) & dex \\
e\_LOG\_G & F.4.2 & Error on LOG\_G & dex \\
LOG\_Y & F5.2 & Log helium abundance $n({\rm He})/n({\rm H})$ & dex \\
e\_LOG\_Y & F5.2 & Error on LOG\_Y & dex \\
PARAMS\_REF & A20 & Reference for atmospheric parameters (Bibcode) &  \\
EB-V & F6.4 & Instellar reddening E(B-V) & mag \\
e\_EB-V & F6.4 & Error on EB-V & mag \\
AV & F6.4 & Instellar extinction A$_{\rm V}$ & mag \\
FUV\_GALEX & F6.3 & GALEX FUV-band magnitude & mag \\
e\_FUV\_GALEX & F6.3 & Error on FUV\_GALEX & mag \\
NUV\_GALEX & F6.3 & GALEX NUV-band magnitude & mag \\
e\_NUV\_GALEX & F6.3 & Error on NUV\_GALEX & mag \\
FUV\_GALEX\_CORR & F6.3 & GALEX FUV-band magnitude corrected (Camarota \& Holberg \cite{camarota14}) & mag \\
NUV\_GALEX\_CORR & F6.3 & GALEX NUV-band magnitude corrected (Camarota \& Holberg \cite{camarota14}) & mag \\
F\_GSC & F6.3 & Guide Star Catalogue R$_{\rm F}$-band magnitude & mag \\
e\_F\_GSC & F6.3 & Error on F\_GSC & mag \\
Bj\_GSC & F6.3 & Guide Star Catalogue B$_{\rm j}$-band magnitude & mag \\
e\_Bj\_GSC & F6.3 & Error on Bj\_GSC & mag \\
V\_GSC & F6.3 & Guide Star Catalogue V-band magnitude & mag \\
e\_V\_GSC & F6.3 & Error on V\_GSC & mag \\
N\_GSC & F6.3 & Guide Star Catalogue I$_{\rm N}$-band magnitude & mag \\
e\_N\_GSC & F6.3 & Error on N\_GSC & mag \\
B\_GSC & F6.3 & Guide Star Catalogue B-band magnitude & mag \\
e\_B\_GSC & F6.3 & Error on B\_GSC & mag \\
V\_APASS & F6.3 & APASS V-band magnitude & mag \\
e\_V\_APASS & F6.3 & Error on V\_APASS & mag \\
B\_APASS & F6.3 & APASS B-band magnitude & mag \\
e\_B\_APASS & F6.3 & Error on V\_APASS & mag \\
g\_APASS & F6.3 & APASS g-band magnitude & mag \\
e\_g\_APASS & F6.3 & Error on g\_APASS & mag \\
r\_APASS & F6.3 & APASS r-band magnitude & mag \\
e\_r\_APASS & F6.3 & Error on r\_APASS & mag \\
i\_APASS & F6.3 & APASS i-band magnitude & mag \\
e\_i\_APASS & F6.3 & Error on i\_APASS & mag \\
u\_SDSS & F6.3 & SDSS u-band magnitude & mag \\
e\_u\_SDSS & F6.3 & Error on u\_SDSS & mag \\
g\_SDSS & F6.3 & SDSS g-band magnitude & mag \\
e\_g\_SDSS & F6.3 & Error on g\_SDSS & mag \\
r\_SDSS & F6.3 & SDSS r-band magnitude & mag \\
e\_r\_SDSS & F6.3 & Error on r\_SDSS & mag \\
i\_SDSS & F6.3 & SDSS i-band magnitude & mag \\
e\_i\_SDSS & F6.3 & Error on i\_SDSS & mag \\
z\_SDSS & F6.3 & SDSS z-band magnitude & mag \\
e\_z\_SDSS & F6.3 & Error on z\_SDSS & mag \\
u\_VST & F6.3 & VST surveys (ATLAS, KiDS) u-band magnitude & mag \\
e\_u\_VST & F6.3 & Error on u\_VST & mag \\
g\_VST & F6.3 & VST surveys (ATLAS, KiDS) g-band magnitude & mag \\
e\_g\_VST & F6.3 & Error on g\_VST & mag \\
r\_VST & F6.3 & VST surveys (ATLAS, KiDS) r-band magnitude & mag \\
e\_r\_VST & F6.3 & Error on r\_VST & mag \\
i\_VST & F6.3 & VST surveys (ATLAS, KiDS) i-band magnitude & mag \\
e\_i\_VST & F6.3 & Error on i\_VST & mag \\
z\_VST & F6.3 & VST surveys (ATLAS, KiDS) z-band magnitude & mag \\
e\_z\_VST & F6.3 & Error on z\_VST & mag \\
J\_2MASS & F6.3 & 2MASS J-band magnitude & mag \\
e\_J\_2MASS & F6.3 & Error on J\_2MASS & mag \\
H\_2MASS & F6.3 & 2MASS H-band magnitude & mag \\
e\_H\_2MASS & F6.3 & Error on H\_2MASS & mag \\
K\_2MASS & F6.3 & 2MASS K-band magnitude & mag \\
e\_K\_2MASS & F6.3 & Error on K\_2MASS & mag \\
Y\_UKIDSS & F6.3 & UKIDSS Y-band magnitude & mag \\
e\_Y\_UKIDSS & F6.3 & Error on Y\_UKIDSS & mag \\
J\_UKIDSS & F6.3 & UKIDSS J-band magnitude & mag \\
e\_J\_UKIDSS & F6.3 & Error on J\_UKIDSS & mag \\
H\_UKIDSS & F6.3 & UKIDSS H-band magnitude & mag \\
e\_H\_UKIDSS & F6.3 & Error on H\_UKIDSS & mag \\
K\_UKIDSS & F6.3 & UKIDSS K-band magnitude & mag \\
e\_K\_UKIDSS & F6.3 & Error on K\_UKIDSS & mag \\
Z\_VISTA & F6.3 & VISTA surveys (VHS, VIKING) Z-band magnitude & mag \\
e\_Z\_VISTA & F6.3 & Error on Z\_VISTA & mag \\
Y\_VISTA & F6.3 & VISTA surveys (VHS, VIKING) Y-band magnitude & mag \\
e\_Y\_VISTA & F6.3 & Error on Y\_VISTA & mag \\
J\_VISTA & F6.3 & VISTA surveys (VHS, VIKING) J-band magnitude & mag \\
e\_J\_VISTA & F6.3 & Error on J\_VISTA & mag \\
H\_VISTA & F6.3 & VISTA surveys (VHS, VIKING) H-band magnitude & mag \\
e\_H\_VISTA & F6.3 & Error on H\_VISTA & mag \\
Ks\_VISTA & F6.3 & VISTA surveys (VHS, VIKING) Ks-band magnitude & mag \\
e\_Ks\_VISTA & F6.3 & Error on Ks\_VISTA & mag \\
W1 & F6.3 & WISE W1-band magnitude & mag \\
e\_W1 & F6.3 & Error on W1 & mag \\
W2 & F6.3 & WISE W2-band magnitude & mag \\
e\_W2 & F6.3 & Error on W2 & mag \\
W3 & F6.3 & WISE W3-band magnitude & mag \\
e\_W3 & F6.3 & Error on W3 & mag \\
W4 & F6.3 & WISE W4-band magnitude & mag \\
e\_W4 & F6.3 & Error on W4 & mag \\
PM\_RA\_SDSS & F6.1 & SDSS proper motion $\mu_{\rm \alpha}\cos{\rm \delta}$ & ${\rm mas\,yr^{-1}}$ \\
e\_PM\_RA\_SDSS & F6.1 & Error on PM\_RA\_SDSS & ${\rm mas\,yr^{-1}}$ \\
PM\_DEC\_SDSS & F6.1 & SDSS proper motion $\mu_{\rm \delta}$ & ${\rm mas\,yr^{-1}}$ \\
e\_PM\_DEC\_SDSS & F6.1 & Error on PM\_DEC\_SDSS & ${\rm mas\,yr^{-1}}$ \\
PM\_RA\_UCAC4 & F6.1 & UCAC4 proper motion $\mu_{\rm \alpha}\cos{\rm \delta}$ & ${\rm mas\,yr^{-1}}$ \\
e\_PM\_RA\_UCAC4 & F6.1 & Error on PM\_RA\_UCAC4 & ${\rm mas\,yr^{-1}}$ \\
PM\_DEC\_UCAC4 & F6.1 & UCAC4 proper motion $\mu_{\rm \delta}$ & ${\rm mas\,yr^{-1}}$ \\
e\_PM\_DEC\_UCAC4 & F6.1 & Error on PM\_DEC\_UCAC4 & ${\rm mas\,yr^{-1}}$ \\
PM\_RA\_PPMXL & F6.1 & PPMXL proper motion $\mu_{\rm \alpha}\cos{\rm \delta}$ & ${\rm mas\,yr^{-1}}$ \\
e\_PM\_RA\_PPMXL & F6.1 & Error on PM\_RA\_PPMXL & ${\rm mas\,yr^{-1}}$ \\
PM\_DEC\_PPMXL & F6.1 & PPMXL proper motion $\mu_{\rm \delta}$ & ${\rm mas\,yr^{-1}}$ \\
e\_PM\_DEC\_PPMXL & F6.1 & Error on PM\_DEC\_PPMXL & ${\rm mas\,yr^{-1}}$ \\
PM\_RA\_USNO & F6.1 & USNO B1.0 proper motion $\mu_{\rm \alpha}\cos{\rm \delta}$ & ${\rm mas\,yr^{-1}}$ \\
e\_PM\_RA\_USNO & F6.1 & Error on PM\_RA\_USNO & ${\rm mas\,yr^{-1}}$ \\
PM\_DEC\_USNO & F6.1 & USNO B1.0 proper motion $\mu_{\rm \delta}$ & ${\rm mas\,yr^{-1}}$ \\
e\_PM\_DEC\_USNO & F6.1 & Error on PM\_DEC\_USNO & ${\rm mas\,yr^{-1}}$ \\
PM\_RA\_APOP & F6.1 & APOP proper motion $\mu_{\rm \alpha}\cos{\rm \delta}$ & ${\rm mas\,yr^{-1}}$ \\
e\_PM\_RA\_APOP & F6.1 & Error on PM\_RA\_APOP & ${\rm mas\,yr^{-1}}$ \\
PM\_DEC\_APOP & F6.1 & APOP proper motion $\mu_{\rm \delta}$ & ${\rm mas\,yr^{-1}}$ \\
e\_PM\_DEC\_APOP & F6.1 & Error on PM\_DEC\_APOP & ${\rm mas\,yr^{-1}}$ \\
PM\_RA\_SPM4 & F6.1 & SPM4 proper motion $\mu_{\rm \alpha}\cos{\rm \delta}$ & ${\rm mas\,yr^{-1}}$ \\
e\_PM\_RA\_SPM4 & F6.1 & Error on PM\_RA\_SPM4 & ${\rm mas\,yr^{-1}}$ \\
PM\_DEC\_SPM4 & F6.1 & SPM4 proper motion $\mu_{\rm \delta}$ & ${\rm mas\,yr^{-1}}$ \\
e\_PM\_DEC\_SPM4 & F6.1 & Error on PM\_DEC\_SPM4 & ${\rm mas\,yr^{-1}}$ \\
LC\_CRTS & A30 & CRTS light curve properties &  \\
LC\_SWASP & A30 & SWASP light curve properties &  \\
\noalign{\smallskip}
\hline\hline
\end{longtable}
}

\end{document}